# Precision of silicon oxynitride refractive-index profile retrieval using optical characterization

Vít Kanclíř*, Jan Václavík and Karel Žídek**

*Regional Centre for Special Optics and Optoelectronic Systems (TOPTEC), Institute of Plasma Physics, Academy of Sciences of the Czech Republic, Za Slovankou 1782/3, 182 00 Prague 8, Czech Republic*

ABSTRACT

Layers with gradient refractive-index profile are an attractive alternative to conventional homogeneous stack coatings. However, the optical characterization and monitoring of the graded refractive-index profile is a complex issue, which has been typically solved by using a simplified model of mixed materials. Although this approach provides a solution to the problem, the precision, which can be expected from optical characterization of the refractive index gradient, remains unclear. In this work, we study optical characterization of $SiO_xN_y$ layers deposited via reactive dual ion beam sputtering. To characterize the deposited layers, we use several methods including reflectance, and transmittance spectra at a broad range of incident angles, together with spectral ellipsometry. All the data were simultaneously fitted with a general profile of refractive index. The expected profile used in our fit was based on characterization of $SiO_xN_y$ layers with a varying stoichiometry. By altering of the profile, we discussed sensitivity of such alternation on fit quality and we studied ambiguity of merit-function minimization. We demonstrate that while the scanning of particular parameters of the profile can be seemingly very precise, we obtain a very good agreement between the experimental data and model for a broad range of gradient shapes, where the refractive-index value on major part of the profile can differ as much as 0.02 from the mean value.

## 1. Introduction

Optical coatings consisting of a stack of thin dielectric layers are extensively employed to improve characteristics of optical elements, most commonly by adjusting their reflectance and transmittance. The desired optical response is typically attained by a deposition of alternating materials with a low and a high refractive index. Thicknesses of the layers of each of the materials determine the resulting properties of the stack. For several decades, there has also been employed an alternative approach, where the refractive index within the coating is changing gradually. This approach features several advantages, such as the absence of interfaces between the high and the low refractive-index films which can improve the laser-induced damage threshold [1, 2]. In case of rugate filters, which have graded refractive-index profile, their stop bands are not surrounded by sidelobes and produce no harmonic stop bands, in contrast to quarter-wave stack of homogeneous layers [3].

A material which is commonly used in deposition of gradient layers is silicon oxynitrid ($SiO_xN_y$). Its advantages are transparency in VIS and NIR spectra, together with wide range of refractive index varying between 2.063 for $Si_3N_4$ [4] and 1.468 for $SiO_2$ [5] at the wavelength 500 nm. A variety of techniques can be used for $SiO_xN_y$ deposition, including plasma enhanced chemical vapour deposition [6], ion assisted deposition [7], magnetron sputtering [8], or dual ion beam sputtering (DIBS) [9–12].

In order to attain a desired optical response, it is necessary to control and measure the refractive-index gradient. This is commonly characterized by two non-destructive methods: spectral ellipsometry (SE) and spectrophotometry, i.e. reflectance (R) and transmittance (T) measurements. Both methods are indirect, thus a proper model is needed to retrieve the refractive and extinction index. Typical problem of this method is the ambiguity of solutions, which arises even for the standard layers. The problem can be partly resolved by using measurements of SE, T and R taken at several angles [13, 14].

*Corresponding author
**Principal corresponding author
✉ kanclir@ipp.cas.cz (V. Kanclíř); zidek@ipp.cas.cz (K. Žídek)
ORCID(s): 0000-0001-5422-231X (V. Kanclíř); 0000-0002-3275-2579 (K. Žídek)

The characterization of the graded-index films is even more complex problem as it is necessary to characterize the gradually varying profile. This is typically solved by dividing the layer to several homogeneous sublayers, where their number should be high enough to ensure that the thickness of the sublayers is thinner than quarter of the smallest wavelength in the calculations [15]. A number of groups have already applied this approach using either SE data [6, 7, 15–17], or both SE and spectrophotometric data [12, 13]. Most of the groups used effective medium approximation (EMA) developed by Bruggeman [18] to model the refractive-index profile, where the profile is obtained from the knowledge of relative volume fraction between $SiO_2$ and $Si_3N_4$. Nevertheless, this method strongly depends on accuracy of outer indices and on the accuracy of the volume ratio. At the same time, EMA is a model for composite materials, and thus it is not physical for $SiO_xN_y$ although it is functioning as a reasonable approximation [19]. As a different approach, Tonova et al. [17] implemented a method to reconstruct a general refractive-index profile. She simulated application of the Newton–Kantorovitch algorithm on ellipsometric data measured at multiple angles of incidence.

An important issue connected with the measurements of gradient refractive-index profile consists in the fact that there is no reliable method able to set the true profile. For this reason, it is very problematic to discuss the reconstruction precision, beside the use of synthetic data [17]. In all the listed EMA-based articles, the precision of the attained gradient refractive-index model was not discussed. Hence, it remains unclear what is the actual precision attainable by using commonly used optical characterization of a gradient thin film.

In this paper, we provide a detailed experimental study of optical characterization of an arbitrary gradient refractive index, where we focus on precision of the gradient shape reconstruction. Namely we employed DIBS-deposited $SiO_xN_y$ layers, which were characterized by using a broad set of commonly used methods: SE measurements and spectrophotometry measured at multiple incident angles. We attained an initial estimate of a refractive-index profile from the characterization of individual homogeneous thin films with a distinct oxygen and nitrogen stoichiometry. Then we used the same parameters with a varying stoichiometry to deposit a gradient thin layer and we measured its optical response.

The goal of our work was to alter the expected gradient profile to attain the best agreement between the experimental and the calculated optical response. We focused on the sensitivity of the agreement towards variation in the profile offset, addition or subtraction of a quadratic function, or a random subtle modification of the initial model. While the controlled modification of the gradient refractive-index profile might suggest that the profile can be determined with a very high precision, we demonstrate that the experimental data can be reproduced by a broad set of profiles with the refractive index varying as much as 0.02 around the central value. Therefore, our work provides an insight into the precision which can be expected from refractive-index profile retrieval based on an optical characterization of a general gradient refractive-index thin film.

## 2. Experimental details

We deposited $SiO_xN_y$ thin layers by DIBS apparatus described in [20]. DIBS is employed in this paper for enhanced control of stoichiometry and for qualities of deposited layers i.e. density, adhesion, nucleation etc. [21] The primary ion source sputtered silicon from a target, and the assistant ion source generated reactions of oxygen and nitrogen with the sputtered silicon atoms to form $SiO_xN_y$. The beam voltage and beam current of the primary ion source was set to 600 V and 108 mA, respectively.

The assistant ion source parameters were set to 120 V for discharge voltage and 0.6 A for discharge current. Other parameters of the assistant ion source are representing gas flow. Flow of nitrogen was set to 49 sccm, and flow of oxygen varied within 0 and 3 sccm (see below).

We used plane-parallel N-BK7 as a substrate for all the depositions, except the deposition of homogeneous layers with flow of oxygen 2.5 and 3 sccm. For those, plane-parallel N-SF10 was used because deposited refractive index of the layers was too similar to the one of N-BK7, and lack of contrast would not allow reliable refractive index retrieval.

Both transmittance T and reflectance R spectra were measured within wavelengths 380 and 980 nm by EssenOptics Photon RT spectrometer. The measurements were carried out for the incident angles 4°, 8°, 20°, 30°, 40°, 50°, 60°, and 70°, where the angle of 4° was used only for the measurement of transmittance. All the spectra were measured for both p- and s-polarization.

The ellipsometry measurements were carried out via Sentech SE850 with micro-spots. We measured visible range of wavelengths between 280 nm and 850 nm under the incident angle of 70°.

**Table 1**
Retrieved refractive indices of the deposited homogeneous $SiO_xN_y$ layers at 500 nm for a range of oxygen flows $Q$. Degree of the experiment-theory agreement is represented by merit function $f_{mer}$ – see text for details.

| Q(O$_2$) [sccm] | Substrate | d [nm] | n [-] | f$_{mer}$ [-] |
|---|---|---|---|---|
| 0.00 | N-BK7 | 295 | 2.026 | 1.53 |
| 0.25 | N-BK7 | 291 | 1.994 | 1.14 |
| 0.50 | N-BK7 | 299 | 1.857 | 1.90 |
| 0.75 | N-BK7 | 303 | 1.880 | 1.37 |
| 1.00 | N-BK7 | 327 | 1.746 | 1.63 |
| 1.50 | N-BK7 | 329 | 1.682 | 0.87 |
| 2.00 | N-BK7 | 346 | 1.610 | 1.44 |
| 2.50 | N-SF10 | 352 | 1.565 | 1.25 |
| 3.00 | N-SF10 | 357 | 1.543 | 1.48 |

## 3. Results and discussion

### 3.1. Homogeneous layers

The initial goal was to create a reliable model where the deposition parameters of $SiO_xN_y$ are linked to a certain refractive-index dispersion curve. Therefore, we first deposited a set of samples of layers with homogeneous refractive index where each sample had different stoichiometry of $SiO_xN_y$. In other words, we deposited a standard thin film for each sample featuring step-like refractive-index changes over interfaces.

In order to describe optical response of the optical coating, we applied a commonly used transfer-matrix approach described in detail in literature [22]. The transfer-matrix method allows us to calculate the optical response (T,R and SE spectra at several incident angles), provided that we know the complex refractive index of a layer $n_L$, substrate and incident medium, together with the layer thickness. It is worth noting that the listed measurements represent the complete linear optical response of the sample. In order to reproduce the experimental data, we correct the transmittance and reflectance curve for a reflection from the uncoated (backside) substrate surface. The multiple reflections from the substrate backside were neglected.

In order to describe spectral shape of the optical characteristics for incident angle $\theta$ and wavelengths $\lambda$, we used the Tauc-Lorentz model of refractive index [23]. This model is commonly used for wide-band gap dielectric materials. We used two oscillators, where in all the studied cases, one oscillator dominated and the second one accounted for a minor correction (1 % of the amplitude). Parameters of the model were fitted by using a complete dataset acquired for a homogeneous layer of $SiO_xN_y$. The fitting included all the $T$, $R$, $\Psi$ and $\Delta$ curves, and all of them optimized the model parameters with respect to the merit function:

$$f_{mer} = \eta_T \sqrt{\sum_{\theta,\lambda}\left(\tilde{T}_{\theta\lambda} - T_{\theta\lambda}\right)^2} + \eta_R \sqrt{\sum_{\theta,\lambda}\left(\tilde{R}_{\theta\lambda} - R_{\theta\lambda}\right)^2} + \eta_\Delta \sqrt{\sum_{\theta,\lambda}\left(\tilde{\Delta}_{\theta\lambda} - \Delta_{\theta\lambda}\right)^2} + \eta_\Psi \sqrt{\sum_{\theta,\lambda}\left(\tilde{\Psi}_{\theta\lambda} - \Psi_{\theta\lambda}\right)^2}, \quad (1)$$

where tilde marks theoretical spectra made of Tauc-Lorentz model, whereas letters without tilde represent measured spectra. Weight coefficients of transmittance ($\eta_T$), reflectance ($\eta_R$), function $\Psi$ ($\eta_\Psi$) and function $\Delta$ ($\eta_\Delta$) were set to 0.474, 0.474, 0.037 and 0.0046, respectively, to account for the different range and noise level of the T, R and ellipsometric functions. For the given weights, all the types of data contributed comparably to the resulting merit function values.

Fig. 1, upper panels, provides a comparison between the experimentally measured data (blue lines) and the model-based calculated curves (red lines) for selected incident angles and polarization. The applied Tauc-Lorentz model allowed us to achieve a refractive-index profile (Fig. 1, lower panel) with a nearly ideal agreement. The fit was carried out for the whole set of deposited homogeneous layers, where the oxygen flow into the assistant ion source controlled the $SiO_xN_y$ composition – see Table 1 for details. In close agreement with the previously reported values , we observe the refractive index to decrease with the increasing oxygen flow rate, ranging from 2.02 (0 sccm, wavelength 500 nm) up to 1.54 (3 sccm, wavelength 500 nm), where the value approaches the refractive index of silica [7, 8, 19].

By taking into account that the refractive index is expected to gradually and smoothly change with the oxygen ratio, we can fit the attained refractive-index dependence on the oxygen flow with a 4th order polynomial function – see Fig. 2, upper panel. From the deviation of the data points from the fit in Fig. 2, we can estimate that our model leads to an error in the refractive index within approx. 0.03. We reached the same precision when we employed this procedure on $Si_3N_4$ layers with various thicknesses ranging from 250 nm to 1500 nm. We ascribe this inaccuracy to the limitations posed by the double-oscillator Tauc-Lorenz model, where the model is an approximation of a more complex refractive-index dispersion curve of $SiO_xN_y$.

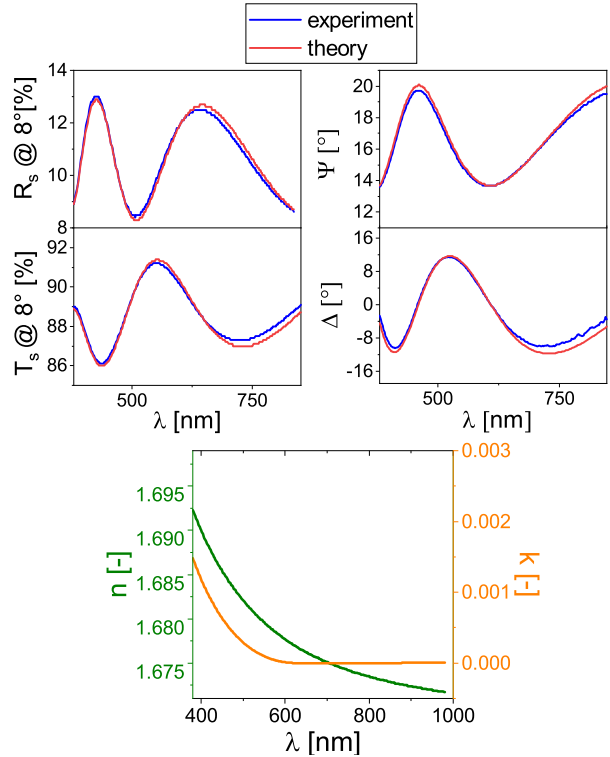

**Figure 1:** Illustration of the homogeneous film characterization (oxygen flow 1.5 sccm). Upper panels: Comparison of experimental (blue lines) and fitted theoretical spectra (red lines) of reflectance $R$, transmittance $T$ ( both s-pol., incident angle 8 deg), and ellipsometric functions $\Delta$ and $\Psi$. (incident angle 70 deg). Lower panel: attained refractive index (green line) and extinction coefficient (orange line) of the $SiO_xN_y$ layer.

### 3.2. Gradient layers

The previous step allowed us to reliably trace changes of the refractive index for a varying oxygen flow in the assistant ion source. As a result, it was possible to attain a layer with a gradient refractive index by a slow variation in the oxygen supply. We deposited the gradient layer with a linear increase of oxygen flow rate in time which leads to the refractive index decreasing with the increasing $z$ – see Fig. 2 upper panel. Note that $z = 0$ corresponds to the substrate surface.

The gradient layers were simulated as a set of 100 thin sublayers with a constant thickness by using the same approach as described in the previous subsection. Owing to the fact that the composition for a depth $z$ can be derived from the deposition procedure, we can assign a refractive index $n(z_i)$ for the sublayers $i$ with a known refractive-index model. The resulting refractive-index gradient $n(z)$ for each wavelength was determined based on spline interpolation of the $n(z_i)$ datapoints – see Fig. 2 upper panel, black line.

Optical response of the gradient layer was fitted by using the same merit function described in Eq. (1). However, the only fitting parameter was the layer thickness and the refractive index was fixed according to the estimated profile depicted in Fig. 2. The resulting merit function of 1.17 was comparable to the ones attained for homogeneous layers. We attained a very good agreement between the experimental data and the theoretical curves – see Fig. 2 lower panel – which we ascribe to the fact that the interpolation of the refractive index partly compensates for the errors in the models of the homogeneous layers.

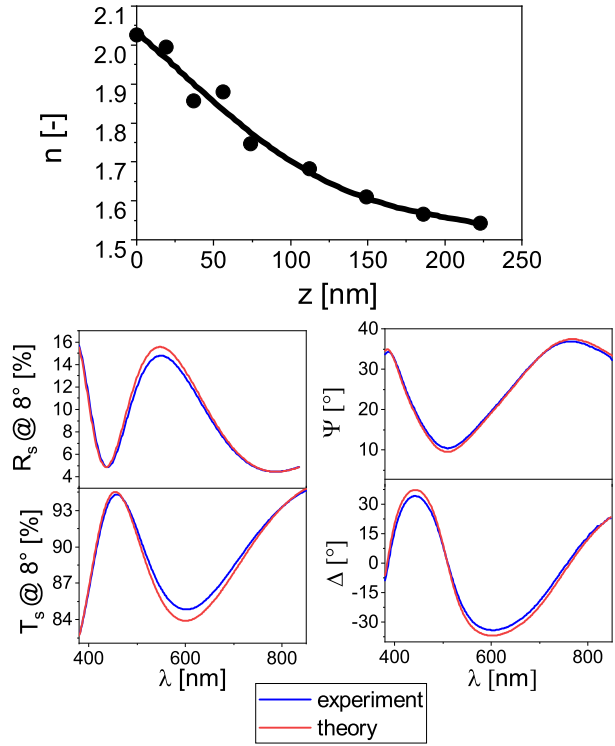

**Figure 2:** Upper panel: Gradient profile of the refractive index at $\lambda = 500$ nm attained by 4th order polynomial fit (line) of the refractive index attained for homogeneous films (circles). Lower panel: Comparison of measured (blue lines) and simulated (red lines) optical response – reflectance, transmittance (both s-pol., incident angle 8 deg), and ellipsometric functions $\Psi$ and $\Delta$ ( incident angle 70 deg).

As a result, we were able to form a reliable initial model of the gradient layer together with its thickness determination. Nevertheless, the ultimate question of interest was to evaluate precision of the estimated profile.

### 3.3. Systematical model variation

Using a free fit of all Tauc-Lorentz parameters for each of the nine concentrations (in total 81 parameters), is prohibitively computationally costly and the substantial number of parameters is likely not to converge to the best result. Therefore, we firstly studied the effect of two main expected imperfections: (i) offset in the gradient refractive index, (ii) change in the overall shape of the gradient.

The offset in a refractive index is one of the problematic issues in material models, since the optical parameters can be often reproduced by using a higher or lower refractive index, which is compensated with a lower or higher layer thickness, respectively. We used the same refractive index as it was used in Fig. 2, and we added an offset of ranging from -5 to 5 percent of the origin refractive-index profile – see Fig. 3 a). For each offset, we fitted the layer thickness and compared the merit functions of the samples – see Fig. 3 c). We observe that the optimum offset was placed around 2 % within the expected value. The effect of the offset on $\Psi$ function around $\lambda = 500$ nm is illustrated in Fig. 3 b).

Secondly, we evaluated the effect of the change in the gradient shape. We altered the original gradient presented in Fig. 2 by adding a quadratic term – see Fig. 3 d). Analogously to the previous case, we fitted the layer thickness in order to achieve the best merit function – see Fig. 3 f).

We observed that the merit function values, i.e., the agreement between the experimental value and our gradient refractive-index model, are highly dependent on both offset and gradient profile bending. This suggests that we can, in principle, determine the gradient refractive-index profile with a very high precision. Optimum offset can be located within 0.25 % precision, implying the refractive-index precision reaching 0.005. We verified that we can optimize the profile curve in arbitrary order of the parameters and we attained closely-lying curves, where the one with the lowest merit function is depicted in Fig. 4 (dashed black line). Therefore, the systematic fitting suggests that we can determine the refractive-index profile with a very high precision. However, as we will show in the next section, the systematical variation of the refractive-index model highly underestimates the actual profile determination inaccuracy.

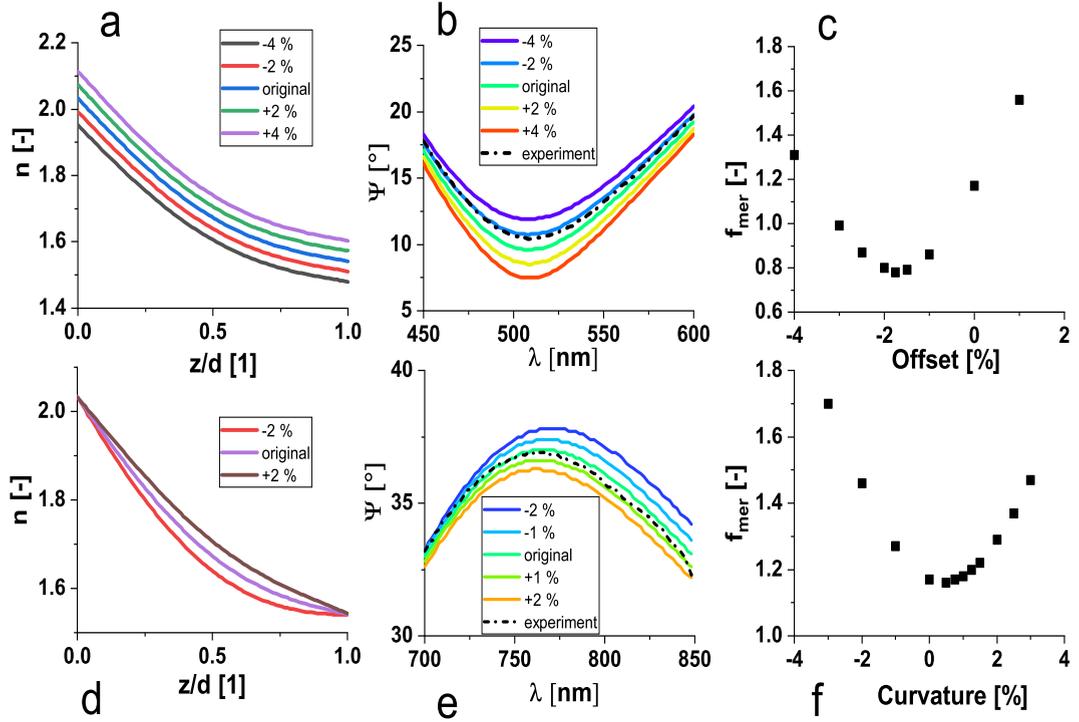

**Figure 3**: Fit evaluation for systematic refractive-index variations by variation of offset (upper panels) and quadratic term addition (lower panels). a,d) Examples of evaluated profiles with a different parameters. b,e) The sensitivity of Ψ function to the shape variation. c,f) A change of the merit-function value with to the shape variation.

### 3.4. Random model variation

To test the actual precision of the optical characterization, we created a simulation, where the refractive indices presented in the upper panel of Fig. 2 (symbols) were altered by a set of random offsets with uniform distribution in the range of ±0.06. For each case, we fitted the refractive-index profile with the 4th order polynomial and carried out the fitting of the layer thickness. By doing 5000 simulations, we attained an extensive set of gradient profiles with minute variations in their shape, which closely followed the original curve, and which featured in some cases the merit function value below the systematically optimized curve.

We selected the randomly modified profiles with merit function below 0.8063, which corresponds to 10 % higher value compared to the best value attained via the systematical gradient variation (0.733) – see colored lines in Fig. 4. The curves with the lowest merit function differ from the systematically attained curve, and we evaluated their spread at several $z$ points. The smallest spread is at $z = 200$ nm and $z = 225$ nm, where the value ranges are 1.559 – 1.577, and 1.537 and 1.555, respectively. More generally, at most points the values differ by 0.02 from the mean value. The spread is the biggest at the edges of the layer – for instance, at $z = 0$, the refractive-index values vary from 1.969 to 2.043. However, this can be ascribed to the polynomial fitting of the data points, which is affected by the random change in the position of the last data point.

The comparison between the curve retrieved by the systematical variation (black dashed line in Fig. 4) with the randomly tested curves (colored lines in Fig. 4) raises a question why the systematical optimization did not converge to a curve located in the center of the best matching profiles of the random simulations. This is caused by the fact that even a minor change in the refractive-index profile highly alters the optimum offset and curvature. As an illustration, only by increasing the refractive index expected at 1 sccm flow of oxygen by +0.02, we attain, by the systematic optimization,

Precision of silicon oxynitride refractive-index profile retrieval using optical characterization a profile (solid black curve in Fig. 4), which lies in the center of the randomly optimized profile. In such case, we also reached a significantly lower merit function value of 0.681. Therefore, the systematically varied parameters, in spite of featuring a sharp optimum value, do not reveal the actual profile precision.

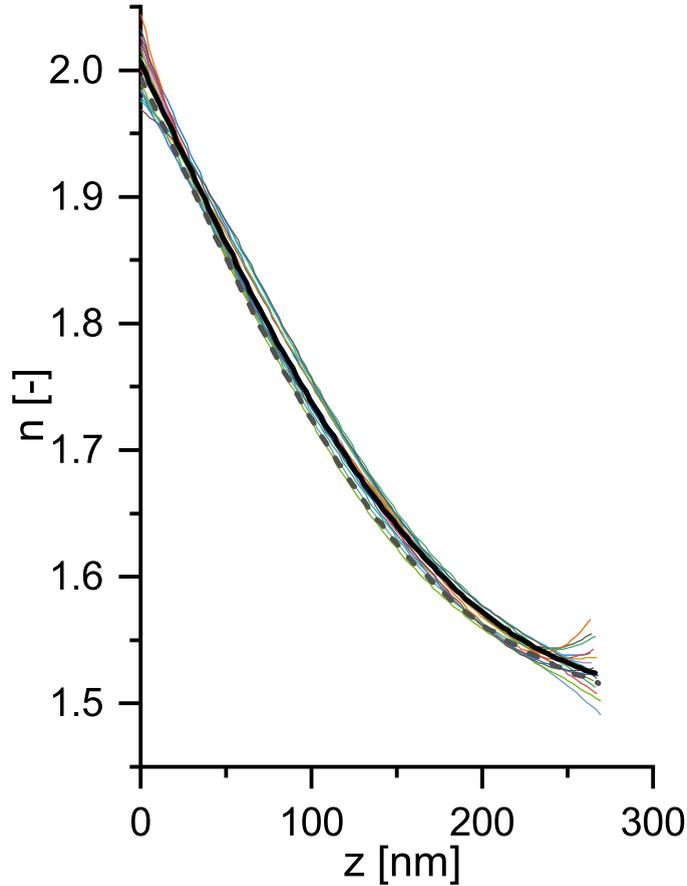

**Figure 4**: Comparison of the refractive-index profiles attained by the random modification of refractive-index datapoints (colored lines) with the systematically optimized ones (black lines – see text for details). Randomly modified curves were selected so that their fitting merit function was lower compared to the systematic optimization (12 curves out of 5000 modifications).

## 4. Conclusions

We carried out a detailed study of optical characterization of $SiO_xN_y$ graded thin films, where we aimed at extracting the refractive-index gradient shape based on the measurements of transmittance, reflectance and ellipsometry. We first attained an estimate of the refractive-index profile based on characterization of $SiO_xN_y$ homogeneous thin films with varying stoichiometry. The estimated profile was varied in order to get the best agreement between the measured and simulated gradient thin film optical response.

We observed that the results might be seemingly very accurate when we scan one particular parameter – for instance, an offset of the gradient. Nevertheless, we can reproduce our experimental data with high precision by using a relatively broad range of gradients with the refractive index varying at most points of the profile by 0.02 from the mean value. We observe that the particular agreement highly varies with subtle modification of the fixed points.

We propose that the precision of the measurement can be improved partly by forming a complex model, where we take into account multiple reflections on the thin layers and substrate. However, the solution of this problem highly depends on a particular experimental device, such as the dimension of the spectrometer's detector and light collection.

Another prominent pathway is to propose a complex specific gradient shape, where the optical response will be more sensitive to the gradient actual shape.

## 5. Acknowledgement

We gratefully acknowledge the financial support of Ministry of Education, Youth and Sports ("Partnership for Excellence in Superprecise Optics," Reg. No. CZ.02.1.01/0.0/0.0/16_026/0008390), and The Czech Academy of Sciences (Strategy AV21, programme 17).